# Leveraging Machine Learning for Early Detection of Lung Diseases

[1]B. Rahmani, [2]H. V. R. Bindela, [2]R. K. R. Gosula, [2]K. C. Yedubati, [2]M. A. Salari, [1]L. Hinyard, [3]P. Norouzzadeh, [4]E. Snir, [1]M. Schoen

[1]Saint Louis University, School of Medicine, Saint Louis, MO

[2]Saint Louis University, Computer Science Department, Saint Louis, MO

[3]Saint Louis University, Professional Studies Department, Saint Louis, MO

[4]Washington University at Saint Louis, Business School, Saint Louis, MO

Corresponding Author:

Bahareh Rahmani, Bahareh.Rahmani@slu.edu

**Abstract**

A combination of traditional image processing methods with advanced neural networks concretes a predictive and preventive healthcare paradigm. This study offers rapid, accurate, and non-invasive diagnostic solutions that can significantly impact patient outcomes, particularly in areas with limited access to radiologists and healthcare resources. In this project, deep learning methods apply in enhancing the diagnosis of respiratory diseases such as COVID-19, lung cancer, and pneumonia from chest x-rays. We trained and validated various neural network models, including CNNs, VGG16, InceptionV3, and EfficientNetB0, with high accuracy, precision, recall, and F1 scores to highlight the models' reliability and potential in real-world diagnostic applications.

## 1. Introduction

The integration of advanced computational methods, especially deep learning, is revolutionizing the field of medical diagnostics. Deep learning methods are especially effective in enhancing image processing. Our project explores the development of machine learning models to enhance the diagnosis of diseases like COVID-19, lung cancer, and pneumonia from chest X-rays, utilizing traditional image processing techniques alongside innovative neural network architectures.

Wang et al. (2020) demonstrated the efficacy of deep learning in diagnosing COVID-19 pneumonia using a novel correlation coefficient, enhancing the precision of chest CT scans. Their model, built on the Xception and Vgg16 frameworks, showed remarkable accuracy in distinguishing COVID-19 cases from other viral pneumonia [1][2].

In the realm of lung cancer, advanced imaging techniques have contributed significantly to early cancer detection, facilitating early interventions that potentially improve patient survival rates [3].

For pneumonia, Ayan and Ünver (2019) and Muhammad et al. (2021) developed neural network models that utilize deep learning and machine learning techniques to diagnose the disease with greater accuracy, significantly impacting areas with a shortage of radiologists [4][5].

Furthermore, Yu et al. (2021) investigated environmental factors affecting respiratory diseases using deep learning image segmentation and a novel green view index, offering new insights into how vegetation impacts respiratory health [6].

In addition to respiratory conditions, Wang et al. (2023) introduced a Convolutional Neural Network (CNN)-based method for detecting dental cavities, showcasing the potential of machine learning to transform dental health care by providing efficient, non-invasive diagnostics [7].

Lastly, Wei et al. (2023) employed machine learning classifiers to analyze COVID-19 distribution patterns in China, using K-Means clustering and regression analysis to explore the demographic and geographic nuances of infection clusters, thereby enhancing pandemic management strategies [8].

Furthermore, recent studies have demonstrated the effectiveness of sequential, functional, and transfer models in detecting lung diseases such as pneumonia, tuberculosis, and lung cancer from both X-ray and CT images. These models have achieved high accuracy levels, supporting their potential use in early diagnosis, particularly in settings where trained professionals are scarce [9].

Additionally, recent work has explored the benefits of combining chest X-ray and CT images to improve diagnostic accuracy for COVID-19, pneumonia, and lung cancer. By using both imaging types, deep learning models can accurately detect disease-specific features at different stages, addressing diagnostic challenges posed by these overlapping conditions [10].

VGG16-based models, for example, have shown high accuracy in detecting and classifying multiple lung diseases, including COVID-19, pneumonia, and pneumothorax, with performance metrics ranging between 93% and 100% [11]. Moreover, VGG and ResNet deep learning models have demonstrated promising results, achieving accuracy levels up to 99.35% in COVID-19 detection, highlighting their potential for rapid, automatic identification of severe respiratory conditions [12].

Other studies utilizing ResNet50 and DenseNet models have achieved effective classification for COVID-19 diagnosis, with DenseNet models reaching a high validation accuracy of 98.33%, thus providing a reliable solution for early lung disease detection [13]. Additionally, advanced frameworks such as the Multi-modal Classification of Lung Disease and Severity Grading (MCLSG) allow not only for disease identification but also for severity grading, supporting precision medicine by stratifying patients into severity levels and optimizing treatment approaches [14].

Low-cost and efficient approaches like Deep Convolutional Neural Networks (DCNN) have also proven effective for widespread screening, especially in resource-limited areas, making them ideal for the early detection of COVID-19, pneumonia, and tuberculosis [15]. Furthermore, systematic reviews highlight those pre-trained architectures such as ResNet, VGG, and DenseNet are among the most used models for lung disease detection; combining these networks with robust classifiers can enhance diagnostic performance and accuracy [16].

Hybrid deep learning algorithms, such as the Hybrid Deep Learning Algorithm (HDLA) framework, combine CNN-based feature extraction with classifiers like SVM and AdaBoost to optimize classification accuracy and computational efficiency for lung disease detection [17]. Another recent framework, VDSNet, integrates VGG with spatial transformer networks (STN) and data augmentation to improve performance on rotated or tilted X-ray images, achieving higher accuracy and making lung disease detection more accessible for medical professionals [18].

These studies collectively advance the capabilities of medical diagnostics, providing a foundation for the continued development and refinement of predictive models. The combination of traditional image processing methods with advanced neural networks not only enhances disease detection accuracy but also paves the way for a predictive and preventive healthcare paradigm.

## 2. Data Description

### 2.1 Covid Dataset

The COVID dataset is a crucial resource comprising 317 JPEG images, meticulously categorized into 'COVID', 'Normal', and 'Viral Pneumonia'. Organized into 'train' and 'test' branches, this dataset ensures streamlined access and efficient navigation, facilitating both model training and validation. Originating from Kaggle, this collection provides a foundational platform for developing sophisticated models aimed at distinguishing these respiratory conditions effectively. Figure 1 presents sample images from each category, illustrating the distinct visual features that help models differentiate between COVID-19, normal lung conditions, and viral pneumonia. These sample images reflect the dataset's capacity to represent unique patterns across categories, aiding in model training.

The 'train' branch includes 111 images classified as 'COVID', 70 as 'Normal', and 70 as 'Viral Pneumonia', providing a rich variety of data to enhance model learning capabilities through exposure to diverse patterns and features inherent in different conditions. Conversely, the 'test' branch is critical for evaluating model robustness and accuracy on previously unseen data, comprising 26 'COVID' images, 20 'Normal', and 20 'Viral Pneumonia'. This setup ensures that the models can be rigorously tested to assess their performance accurately.

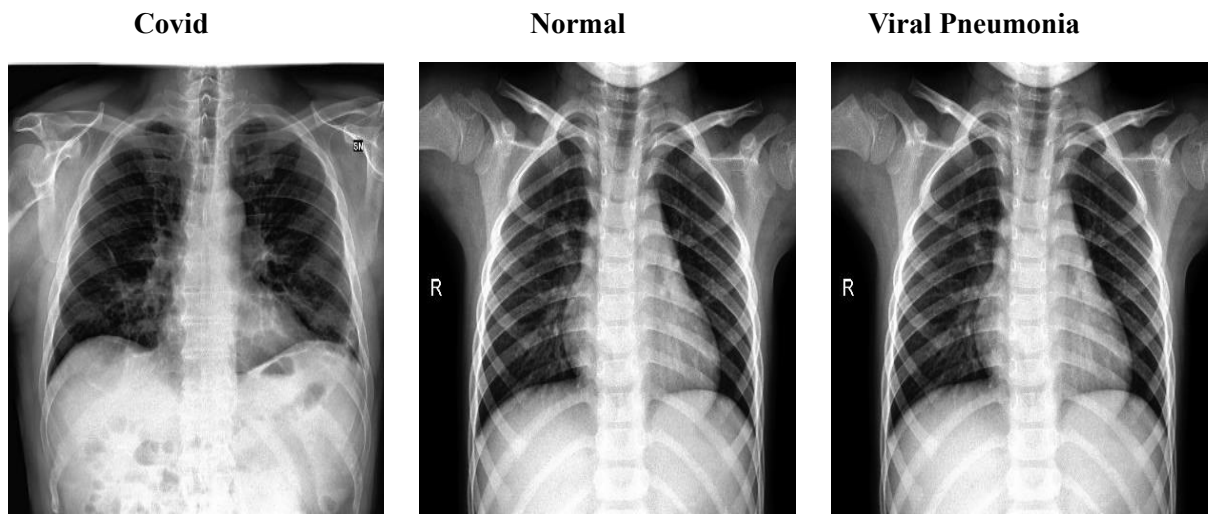

Figure 1: Sample Images from COVID Dataset

### 2.2 Pneumonia Dataset

The Pneumonia dataset comprises 5,860 chest X-ray JPEG images, categorized into 'Pneumonia' and 'Normal' conditions, derived from pediatric patients aged one to five years at the Guangzhou Women and Children's Medical Center. This dataset has undergone extensive quality control, with each image being rigorously screened for clarity and subsequently diagnosed by two expert physicians, with additional evaluations by a third expert for the validation subset. The images are organized into 'train', 'test', and 'valid' folders to facilitate efficient data access and model training. Figure 2 provides representative images from the dataset, illustrating the visual differences between 'Pneumonia' and 'Normal' conditions. These samples highlight the dataset's clarity and diagnostic quality, which are crucial for enhancing model accuracy. The training set includes 5,214 images (1,340 Normal, 3,874 Pneumonia), the testing set comprises 624 images (234 Normal, 390 Pneumonia), and the validation set contains 16 images evenly divided between Normal and Pneumonia (8 each). This structured organization ensures the dataset's suitability for developing robust diagnostic models, enabling detailed study and effective management of pediatric pneumonia through advanced machine learning techniques.

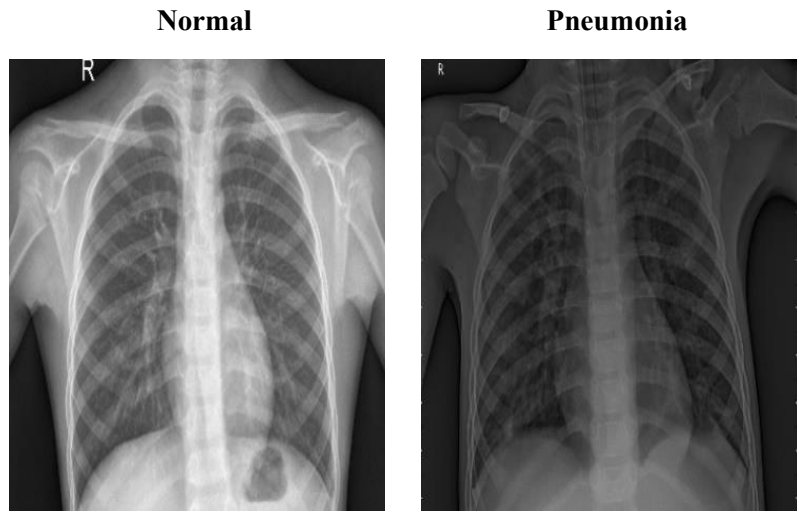

**Normal**     **Pneumonia**

Figure 2: Sample Images from Pneumonia Dataset

## 2.3 Lung Cancer Dataset

The Lung Cancer dataset provides a comprehensive examination of chest conditions through images stored in both JPEG and PNG formats, classifying them into Adenocarcinoma, Large cell carcinoma, Squamous cell carcinoma, and Normal. This dataset is organized into 'test', 'train', and 'valid' folders, partitioned to represent 20%, 70%, and 10% of the dataset, respectively. Figure 3 displays representative images from each category, illustrating the distinct characteristics of 'Adenocarcinoma', 'Large cell carcinoma', 'Squamous cell carcinoma', and 'Normal' lung conditions. These images serve as key visual references, supporting model training to accurately classify each type. The 'test' set comprises 315 images with 54 Normal, 120 Adenocarcinoma, 51 Large cell carcinoma, and 90 Squamous cell carcinomas. The 'train' set includes 613 images featuring 148 Normal, 195 Adenocarcinoma, 115 Large cell carcinoma, and 155 Squamous cell carcinomas, while the 'validation' set contains 72 images with 13 Normal, 23 Adenocarcinoma, 21 Large cell carcinoma, and 15 Squamous cell carcinomas. Adenocarcinoma, the most prevalent form of lung cancer, constitutes approximately 30% of lung cancer cases and 40% of non-small cell lung cancer (NSCLC) cases, typically manifesting in the outer regions of the lungs. Large cell carcinoma, known for its rapid growth, accounts for 10-15% of NSCLC cases and can appear anywhere in the lung. Squamous cell carcinoma, making up about 30% of NSCLC, is primarily found in the central regions of the lung and is often linked to smoking. This dataset's detailed categorization and high-quality imaging make it an invaluable resource for developing diagnostic models that distinguish between different types of lung cancer, enhancing our understanding and management of this complex disease.

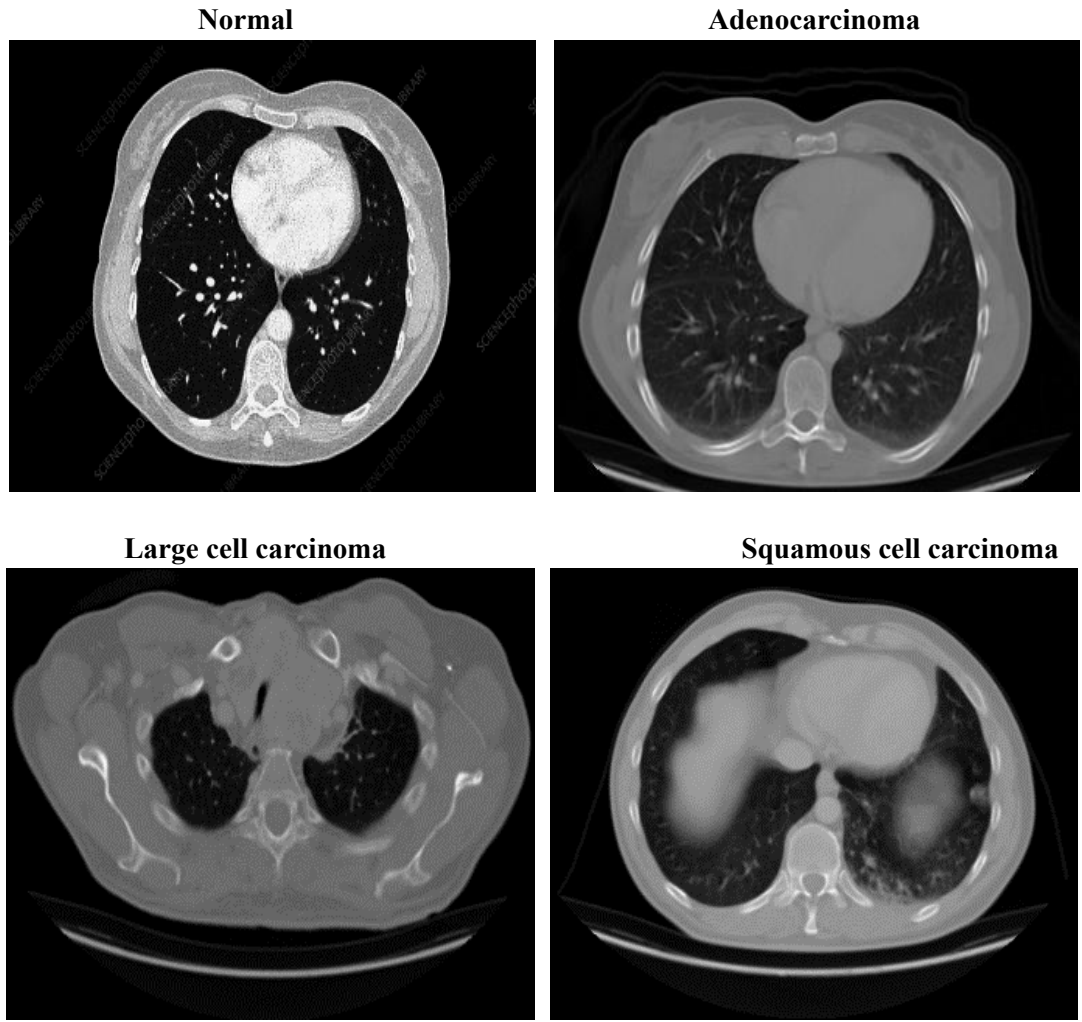

Figure 3: Sample Images from Lung Cancer Dataset

These datasets, all curated from Kaggle, provide critical insights and robust platforms for the development of machine-learning models aimed at enhancing the diagnosis of respiratory diseases and conditions. They represent a significant component of our study, enabling the training and validation of our diagnostic models.

## 3. Methodology

### 3.1 Pneumonia Classification

To address the challenge of accurately classifying chest X-rays as Normal or Pneumonia, we implemented a CNN. This architecture excels in image recognition tasks by efficiently extracting significant features from visual data, making it ideal for detailed medical imaging analysis. Figure 4 shows sample X-ray images from the dataset, highlighting Normal and Pneumonia cases. These images reflect the complexity of visual features that CNN learns to differentiate, aiding in robust classification of pulmonary conditions.

#### 3.1.1 Model Architecture

Our CNN consists of multiple convolutional layers, each followed by pooling and dropout layers to prevent overfitting, and fully connected layers leading to a classification output. We utilize the ReLU activation function in convolutional layers to introduce non-linearities, which are crucial for learning complex patterns in the image data.

### 3.1.2 Data Preparation

The dataset comprises pre-labeled X-ray images categorized as Normal or Pneumonia. We standardized the images to uniform dimensions and normalized them to ensure consistent data scaling across inputs. Data augmentation techniques such as rotations and horizontal flips were implemented to enhance the model's ability to generalize across new and varied image presentations.

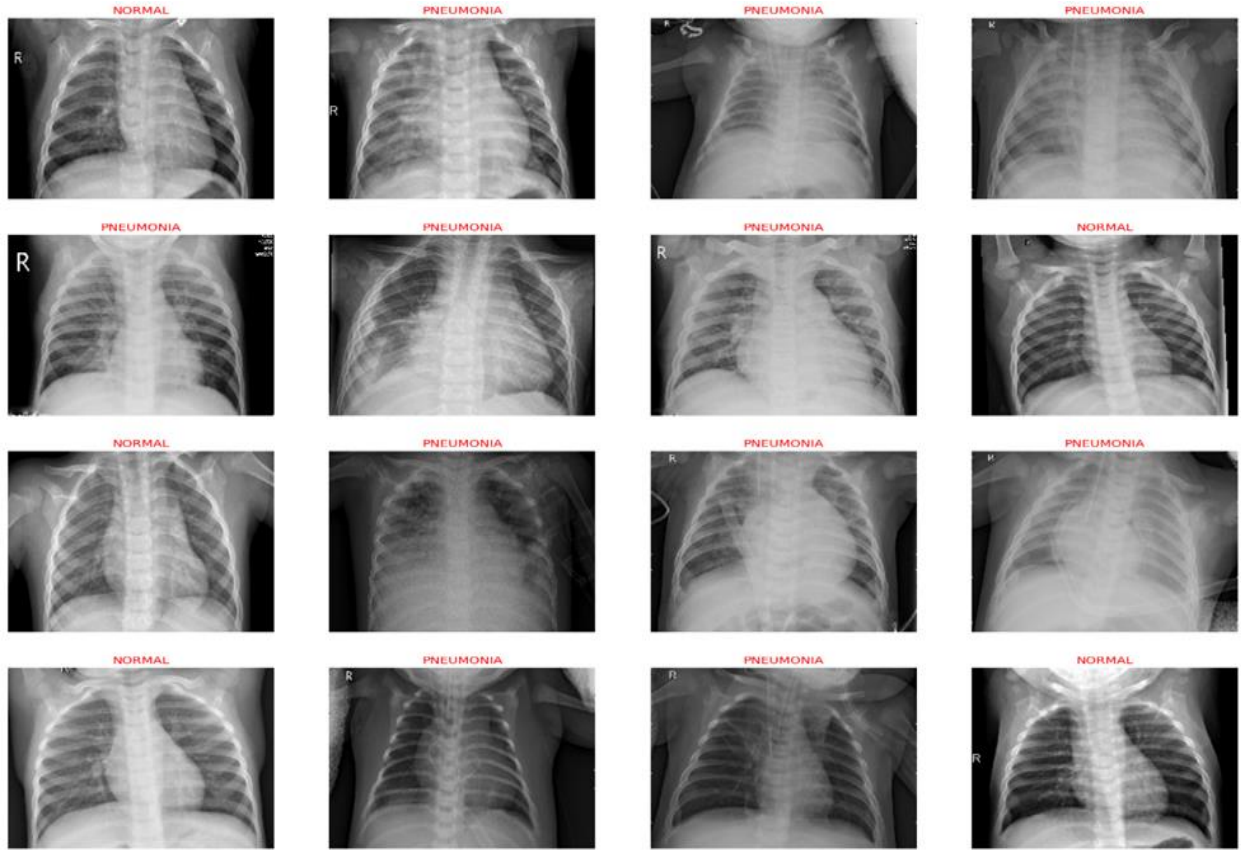

**Figure 4: Sample Chest X-ray Images**

### 3.1.3 Training Process

We randomly allocated 80% of the dataset for training and the remaining 20% for validation, closely monitoring the validation set to prevent overfitting. We chose the Adam optimizer for its adaptive learning rate capabilities, which enhance convergence efficiency during training.

### 3.1.4 Training Metrics

Throughout the training process, we meticulously tracked the training and validation loss and accuracy at each epoch. This monitoring helped us adjust parameters optimally and implement early stopping if the validation loss failed to improve for five consecutive epochs. Figure 5 illustrates the progression of training and validation loss over 50 epochs, demonstrating the point at which early stopping was applied to prevent overfitting and ensure model robustness.

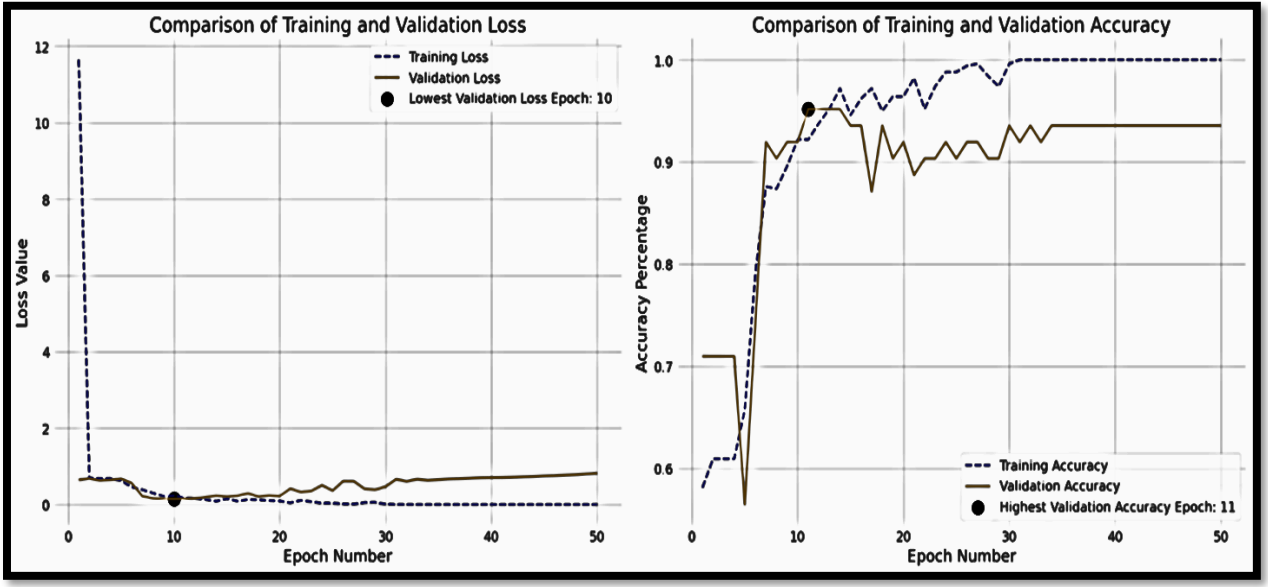

**Figure 5: Training and Validation Loss Over Epochs**

### 3.1.5 Evaluation and Validation

After training, the model was rigorously evaluated on a test set that was not exposed during the training phase to assess its generalizability and real-world applicability. Figure 6 presents the training and validation accuracy of the model across epochs, highlighting the point at which the model reached its peak validation accuracy before leveling off.

**Accuracy:** The model achieved perfect accuracy rates of 100% on both the training set and test set. The validation accuracy was notably high at approximately 93.5%, underscoring the model's effective classification capabilities.

```
Set         |         Loss  | Accuracy (%)
------------+---------------+-------------
Training    | 7.80636e-06   |     100
------------+---------------+-------------
Validation  | 0.816049      |     93.5484
------------+---------------+-------------
Test        | 5.73983e-05   |     100
```

**Figure 6: Training and Validation Accuracy Across Epochs**

**Precision, Recall, and F1-Score:** The model reached high precision and recall for both classes, resulting in F1 scores of 1.00, which indicates a well-balanced capability to accurately identify true positives without significant false positives or negatives. As shown in Figure 7, the model achieves perfect precision, recall, and F1-scores for both Normal and Pneumonia classes, highlighting its accuracy and reliability in classifying chest X-rays. The 'support' column indicates the sample count for each class in the test dataset, confirming robust performance across all metrics.

| precision | recall | f1-score | support |
|---|---|---|---|
| 1 | 1 | 1 | 21 |
| 1 | 1 | 1 | 42 |
| 1 | 1 | 1 | 1 |
| 1 | 1 | 1 | 63 |
| 1 | 1 | 1 | 63 |

**Figure 7: Precision, Recall, and F1-Score for the Classification Model.**

### 3.1.6 Data and Model Evaluation Split

70% of the dataset was used for model training, providing a diverse range of X-ray images for comprehensive learning. 20% of the data was reserved for validation during training, aiding in model adjustments and determining the optimal point to halt training. The remaining 10% served as the test set, utilized solely for final model evaluation to ensure the performance metrics accurately reflect the model's ability to process new, unseen data.

### 3.1.7 Visual Results Analysis

To visually substantiate the model's performance, we employed several graphical representations. Training and validation loss and accuracy plots are crucial for tracking the learning progress and ensuring the model is effectively balanced between generalization and memorization capabilities.

**The Confusion Matrix** provides an in-depth view of the model's accuracy by showing true versus predicted classifications, helping identify specific strengths and weaknesses for further refinement.

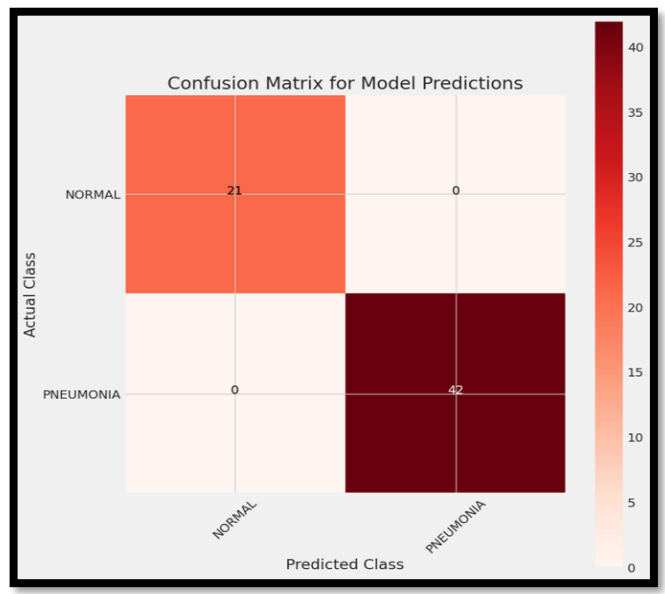

**Figure 8: Confusion Matrix for Model Predictions on Test Data**

Figure 8 displays the confusion matrix, which provides a detailed breakdown of the model's accuracy by showing true versus predicted classifications. This matrix confirms the model's precise performance, with no misclassifications between Normal and Pneumonia cases.

## 3.2 Lung Cancer Classification

To classify chest X-ray images into four categories - Adenocarcinoma, Large Cell Carcinoma, Squamous Cell Carcinoma, and Normal - we implemented deep learning models leveraging pre-trained architectures like VGG16, InceptionV3, and EfficientNetB0. These transfer learning approaches excel at extracting relevant features from medical imaging data, facilitating accurate disease detection. Figure 9 demonstrates this classification process, showing how the model categorizes chest X-ray images into 'Adenocarcinoma', 'Large Cell Carcinoma', 'Squamous Cell Carcinoma', and 'Normal' categories. By utilizing pre-trained architectures for feature extraction, the model achieves heightened accuracy in identifying and distinguishing various lung conditions.

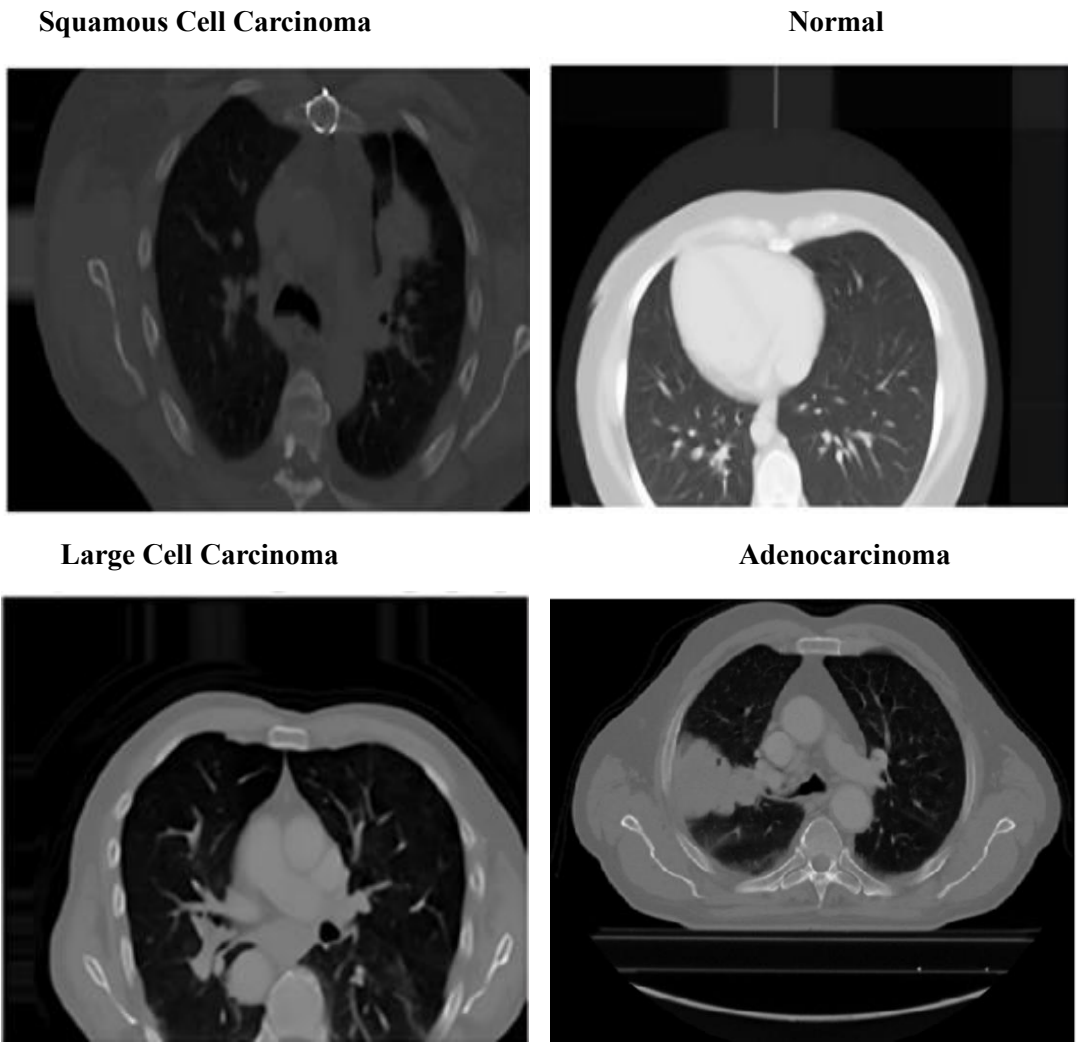

Figure 9: Lung Cancer Classification Using Deep Learning Models

### 3.2.1 Data Preparation

The Lung Cancer dataset comprised JPEG and PNG chest images organized into 'train' (70%), 'valid' (10%), and 'test' (20%) sets. Data preprocessing involved resizing images to 350x350 pixels and normalization for consistent scaling. Augmentation techniques like rotations, flips, zooms, and shifts enriched the training data, enhancing generalization capabilities.

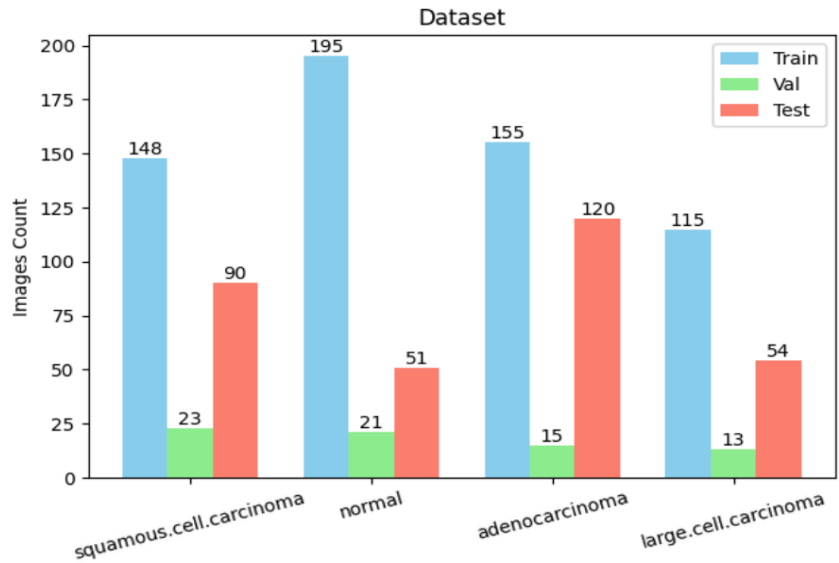

**Figure 10: Dataset Image Distribution for Lung Cancer Classification**

Figure 10 displays the distribution of images across the training, validation, and testing sets for each lung cancer category: Squamous Cell Carcinoma, Normal, Adenocarcinoma, and Large Cell Carcinoma. The blue, green, and red bars represent the training, validation, and test sets, respectively, with image counts labeled for clarity. Figure 11 shows training and validation metrics over 30 epochs, with the left graph tracking accuracy improvements and the right graph displaying loss stabilization. Training and validation losses converge near zero, indicating effective model learning and generalization.

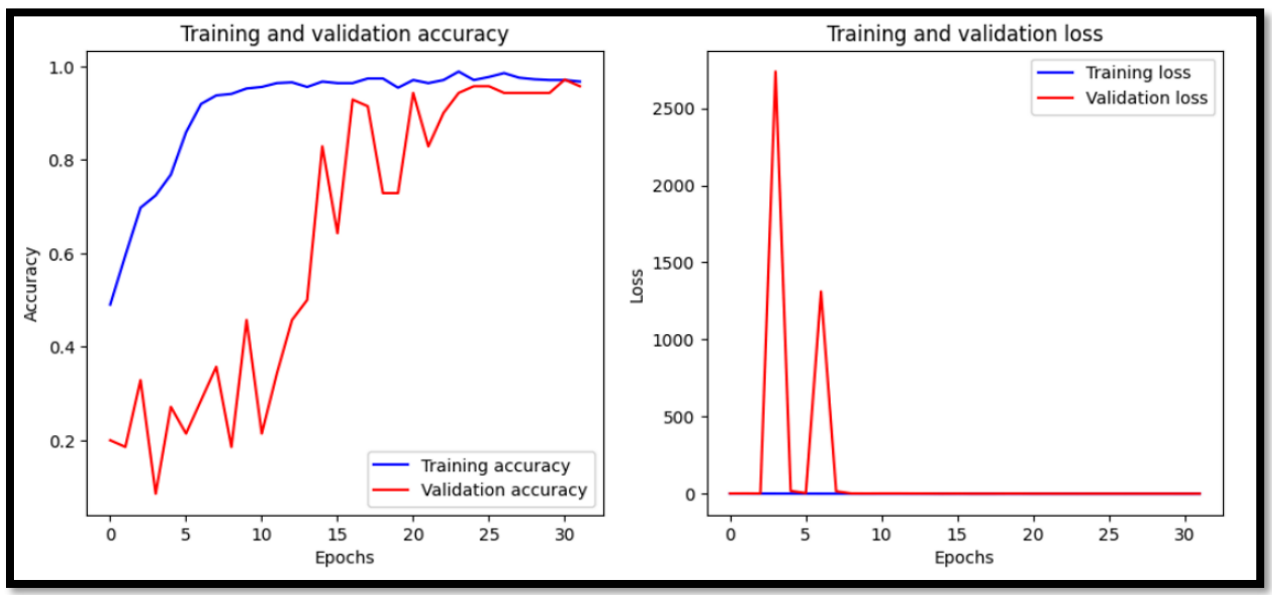

**Figure 11: Training and Validation Metrics**

### 3.2.2 Model Architectures:

**VGG16 Model:** We leveraged the pre-trained VGG16 base without top layers, followed by a Flatten layer, Dropout regularization (0.25), and a Dense layer with 4 neurons and sigmoid activation for classification. The base model's weights were frozen during training.

**InceptionV3 Model:** The pre-trained InceptionV3 base without top layers served as the foundation. Its layers were frozen, and the output was passed through a Flatten layer, Dense layer (1024 neurons, ReLU), Dropout (0.2), and a final Dense layer with 4 neurons and sigmoid activation.

**EfficientNetB0 Model:** We utilized the pre-trained EfficientNetB0 base without top layers, adding a GlobalAveragePooling2D layer, Dropout (0.5), and a Dense layer with 4 neurons and SoftMax activation for classification.

**Training Process:** The models were trained using the Adam optimizer and categorical cross-entropy loss. Validation data monitored overfitting, and callbacks like ModelCheckpoint saved the best models, while ReduceLROnPlateau adjusted learning rates adaptively.

**Evaluation Metrics:** We evaluated the models on the test set, calculating accuracy, loss, precision, recall, and F1 scores. Visual representations like confusion matrices aided performance analysis.

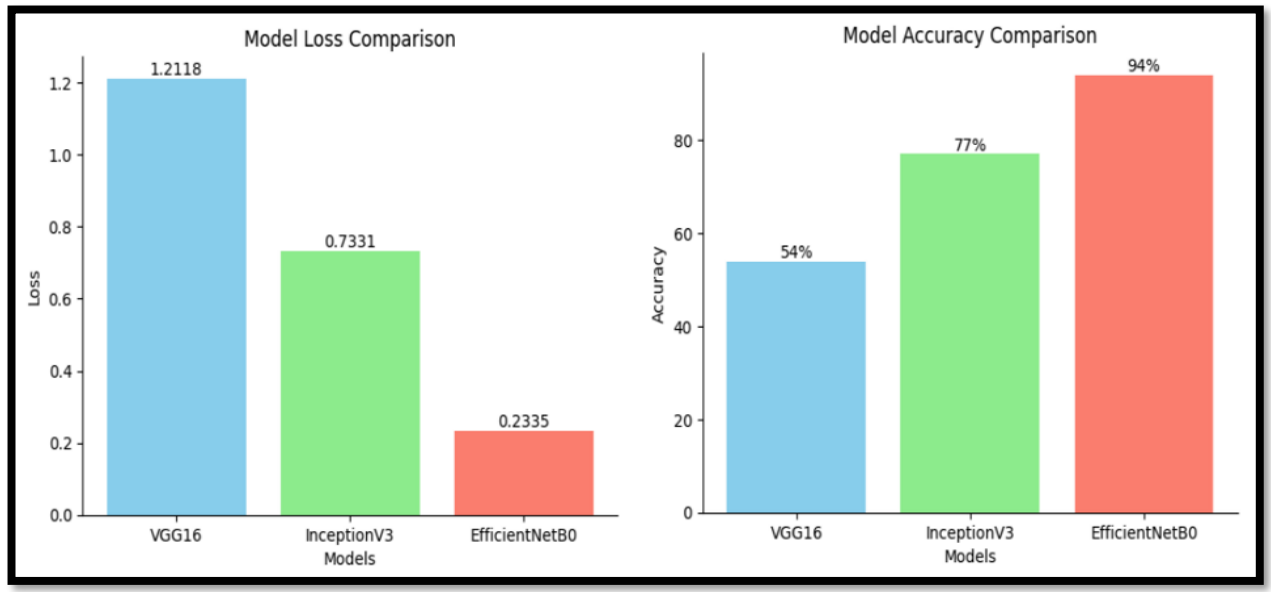

**Figure 12: Model Loss and Accuracy Comparison**

Figure 12 compares model loss and accuracy across the three architectures: VGG16, InceptionV3, and EfficientNetB0. The left graph shows that EfficientNetB0 achieved the lowest loss (0.2335), followed by InceptionV3 and VGG16. The right graph highlights EfficientNetB0's highest accuracy (94%), indicating its superior performance in optimizing both accuracy and loss compared to the other models.

### 3.3 COVID -19 Classification

To address the critical need for rapid and accurate COVID-19 diagnosis using chest X-rays, we employed deep learning techniques, specifically Convolutional Neural Networks (CNNs), due to their robust feature extraction capabilities in image analysis.

### 3.3.1 Model Architecture

To classify chest X-rays into 'COVID', 'Normal', and 'Viral Pneumonia', a Convolutional Neural Network (CNN) was employed. CNNs are particularly effective for image recognition tasks due to their ability to automatically learn spatial hierarchies of features through backpropagation using multiple building blocks such as convolutional layers, pooling layers, and fully connected layers.

Multiple convolutional layers with ReLU (Rectified Linear Unit) activation functions are used to extract features from the input images. ReLU helps introduce non-linearities into the model, allowing it to learn complex patterns. Each convolutional layer is followed by a pooling layer (usually max pooling) to down sample the feature maps, reducing their dimensions and computational complexity while retaining the most important features.

After several convolutional and pooling layers, the network includes fully connected (dense) layers. These layers are used to combine the features extracted by the convolutional layers and make predictions. The last fully connected layer has a number of neurons equal to the number of classes (3 in this case: COVID, Normal, Viral Pneumonia) and uses a SoftMax activation function to produce a probability distribution over the classes.

### 3.3.2. Data Preparation

Data preparation is a crucial step in building a robust and reliable CNN model. It involves transforming raw data into a format that can be efficiently processed by the model and augmenting the data to improve the model's generalization capabilities.

**Standardization:** All chest X-ray images are resized to a standard dimension of 224x224 pixels, ensuring consistency in the input data and making it compatible with the CNN architecture.

**Normalization:** Pixel values of the images are normalized to a range of 0 to 1 by dividing by 255 (the maximum pixel value). Normalization helps in speeding up the training process and achieving better convergence.

**Data Augmentation:** Data augmentation techniques such as random rotations, horizontal flips, and zooming are applied to the training images. This artificially increases the size of the training set and helps the model generalize better by learning from a more diverse set of images.

### 3.3.3. Training Process

The training process for the CNN model was designed to optimize its performance for classifying chest X-rays. The process included careful splitting of the dataset, selection of an appropriate optimizer, and monitoring of training metrics.

**Data Split:** The dataset was split into training and validation sets with an 80-20 ratio. This means that 80% of the data was used for training the model, while the remaining 20% was used for validating its performance. The training set consisted of images from all three classes: COVID, Normal, and Viral Pneumonia.

**Optimizer:** The Adam optimizer was chosen for training the model due to its adaptive learning rate capabilities. Adam combines the advantages of two other extensions of stochastic gradient descent, AdaGrad and RMSProp, which makes it well-suited for dealing with sparse gradients on noisy problems.

**Training Metrics:** Training and validation loss and accuracy were tracked at each epoch. These metrics provided insight into the model's performance and guided adjustments to the training process. Early stopping was implemented to prevent overfitting. If the validation loss did not improve for five consecutive epochs, the training process was halted. This ensured that the model maintained good generalization capabilities.

**Epochs and Batch Size:** The model was trained for a maximum of 50 epochs with a batch size of 32. These hyperparameters were selected based on initial experiments to balance training time and model performance.

### 3.3.4. Evaluation and Validation

After training, the model's performance was rigorously evaluated on a test set. The evaluation metrics included accuracy, precision, recall, and F1-score to provide a comprehensive assessment of the model's classification capabilities.

**Accuracy:** The overall accuracy of the model was calculated as the ratio of correctly predicted instances to the total instances. High accuracy indicates that the model performs well on the test set.

**Classification Report:** The classification report provides a detailed performance analysis, including precision, recall, and F1-score for each class (COVID, Normal, Viral Pneumonia). These metrics are crucial for understanding the model's effectiveness and areas for improvement. Figure 13 displays the classification report with metrics like precision, recall, and F1-score for COVID and Normal classes, showing high precision (1.00 for COVID) and a strong overall accuracy of 0.96, indicating balanced classification performance.

|  | precision | recall | f1-score | support |
|---|---|---|---|---|
| Covid | 1.00 | 0.92 | 0.96 | 26 |
| Normal | 0.91 | 1.00 | 0.95 | 20 |
| Accuracy |  |  | 0.96 | 46 |
| Macro avg | 0.95 | 0.96 | 0.96 | 46 |
| Weighted avg | 0.96 | 0.96 | 0.96 | 46 |

**Figure 13: Classification Report for COVID-19 and Normal Chest X-rays**

**Training and Validation Curves:** Plots of training and validation loss and accuracy over epochs were analyzed to ensure the model was not overfitting. These curves provide visual confirmation of the model's learning progress and help identify any training issues. Figure 14 illustrates the training process, showing an upward trend in accuracy and a decrease in loss over 10 epochs. The stability of these metrics as epochs progress confirms the model's consistent learning and convergence.

In conclusion, our study demonstrates the significant promise of advanced machine learning techniques, particularly deep learning, in enhancing the diagnosis of respiratory diseases such as COVID-19, lung cancer, and pneumonia from chest x-rays. Utilizing datasets from Kaggle, we trained and validated various neural network models, including CNNs, VGG16, InceptionV3, and EfficientNetB0, achieving high accuracy, precision, recall, and F1 scores. These results highlight the models' reliability and potential in real-world diagnostic applications.

The integration of data augmentation and normalization techniques improved the models' robustness, ensuring their generalizability to new data. Our methodology included rigorous training processes and early stopping criteria to prevent overfitting, contributing to the high performance of our models.

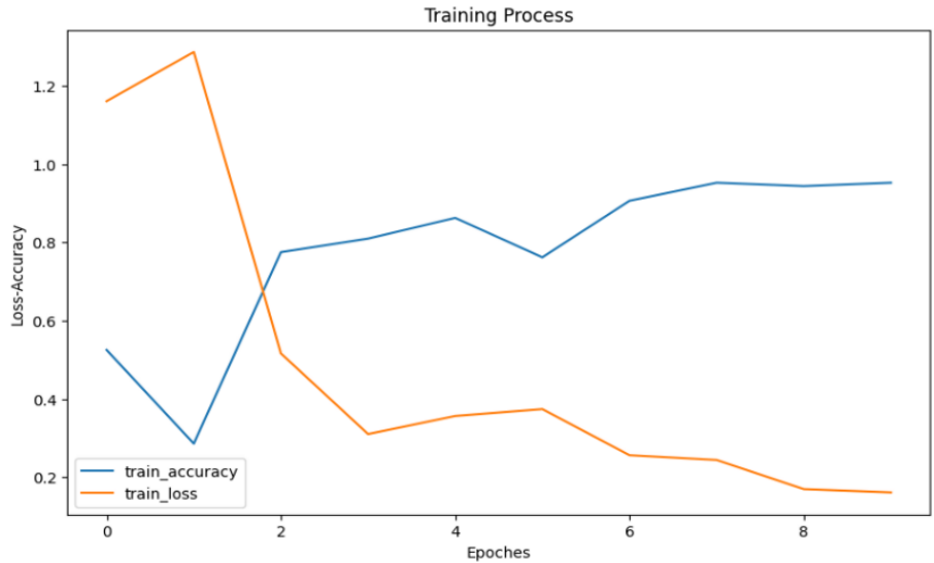

**Figure 14: Training Process; Accuracy and Loss**

Overall, the combination of traditional image processing methods with advanced neural networks paves the way for a predictive and preventive healthcare paradigm. This study provides a foundation for future research, offering rapid, accurate, and non-invasive diagnostic solutions that can significantly impact patient outcomes, particularly in areas with limited access to radiologists and healthcare resources.

**4. Results**

Our models achieved high performance metrics across the board, reflecting their potential applicability in medical diagnostics. For pneumonia classification, the CNN model achieved a validation accuracy of approximately 93.5% and perfect training and test accuracies. For lung cancer, the EfficientNetB0 model outperformed other architectures, achieving a test accuracy of 94%, while for COVID-19 classification, our CNN model reached an accuracy of 96% on the test set. These results confirm the robustness and generalization of our models, making them well-suited for respiratory disease diagnostics.

**5. Discussion**

The high accuracy and reliability of our models suggest that deep learning can play a significant role in non-invasive, rapid diagnostics, especially in under-resourced healthcare settings. Despite achieving good results, further testing in clinical environments would enhance our understanding of these models' practical limitations and strengths. Additionally, future work could explore integrating these models into healthcare workflows, optimizing patient outcomes.

**6. Conclusions**

This study demonstrates the viability of deep learning for diagnosing respiratory diseases, with potential applications for COVID-19, lung cancer, and pneumonia. Utilizing CNNs and pre-trained architectures like VGG16, InceptionV3, and EfficientNetB0, we achieved high accuracy across diverse datasets. The inclusion of data augmentation and normalization further improved robustness, enabling the models to generalize effectively to new data. Our findings support the use of advanced machine learning techniques in predictive and preventive healthcare, particularly in areas lacking access to specialized medical personnel.